\documentclass[12pt]{iopart}
\usepackage{graphicx}
\usepackage{bm}

\newcommand{\be}{\begin{eqnarray}}
\newcommand{\ee}{\end{eqnarray}}
\newcommand{\bmat}{\left(\begin{array}}
\newcommand{\emat}{\end{array}\right)}
\newcommand{\no}{\nonumber}
\newcommand{\la}{\langle}
\newcommand{\ra}{\rangle}

\begin{document}
\title{How fast and robust is the quantum adiabatic passage?}
\author{Kazutaka Takahashi}
\address{Department of Physics, Tokyo Institute of Technology, 
Tokyo 152-8551, Japan}
%\ead{I don't prefer giving my e-mail address here. Just google it!}

\begin{abstract}
We study the assisted adiabatic passage, 
and equivalently the transitionless quantum driving, 
as a quantum brachistochrone trajectory.
The optimal Hamiltonian for given constraints
is constructed from the quantum brachistochrone equation.
We discuss how the adiabatic passage is realized 
as the solution of the equation.
The formulation of the quantum brachistochrone 
is based on the principle of least action.
We utilize it to discuss the stability of 
the adiabatic passage.
\end{abstract}
\maketitle

%%%%%%%%%%%%%%%%%%%%%%%%%%%%%%%%%%%%%%%%%%%%%%%%%%%%%%%%%%%%%%%%%%%%%%%%%%%%
%%%%%%%%%%%%%%%%%%%%%%%%%%%%%%%%%%%%%%%%%%%%%%%%%%%%%%%%%%%%%%%%%%%%%%%%%%%%
\section{Introduction}

Coherent manipulation of quantum systems is 
of fundamental and practical importance and 
has been discussed intensively.
Theoretically, the problem can be stated in a simple question: 
What is the optimal Hamiltonian under a given condition?
For example, we seek the Hamiltonian 
minimizing the traveling time between two states, 
or maximizing the fidelity between the state under evolution 
and a reference state.
Depending on experimental situations, 
we can consider optimizations in many possible ways.
Several optimization methods have been proposed, 
each based on a different philosophy.
The assisted adiabatic passage (AP)~\cite{DR1, DR2}
and the transitionless quantum driving~\cite{Berry}
are known as a nonadiabatic driving following an adiabatic state.
In this method, the counter-diabatic Hamiltonian is introduced 
to make the evolution transitionless.
Several closely related methods have been proposed such as 
shortcuts to adiabaticity or 
the Lewis-Riesenfeld invariant-based engineering~\cite{CRSCGM}
and the fast-forward scaling~\cite{MN1, MN2, MN3}.
These methods have been intensively studied in recent 
years~\cite{MCILR, CLRGM, CM, CTM, IMCTM, Betal, TMRM, ICTMR, BCSM, dCRZ, RCAM, KT, Zetal, Ban, TIMMCGRCM, DRSG, LCRAGM}.

It has been discussed in, e.g., \cite{CTM, TMRM} 
that some of those methods are related 
with each other.
This implies that there exists a common principle behind.
It can be useful to reformulate the problem 
from a different general perspective.
In this paper, we study the AP by using the method of 
the quantum brachistochrone (QB)~\cite{CHKO1, CHKO2, KO}.
For a quantum trajectory, we define an action 
and the optimal Hamiltonian and state are determined from 
the variational principle.
The advantage of this method is in its simplicity and generality.
The QB equation is derived in a general form and 
the system-dependent conditions are implemented in the constraint part.
By using the QB equation, we show that 
the solution of the equation is interpreted as the AP.
It is not trivial what kind of adiabatic state appears as the solution 
and we examine the equation in several examples.

We also study the stability of the AP.
In several examples, the AP was shown to be robust against 
variations of control parameters~\cite{Betal, KT}.
This problem was closely studied in~\cite{RCAM, DRSG, LCRAGM}.
Since the QB is formulated by the variational principle,
we can consider the stability of the solution in a rather general way.
We derive a general stability condition and
examine several examples explicitly.

The paper is organized as follows.
In section~\ref{sec:qb}, we formulate the QB method and 
discuss the relation with the method of the AP.
Then, several examples are studied in section \ref{sec:ex}. 
The stability of the AP is discussed in section \ref{sec:stab}.
Section \ref{sec:summary} is devoted to summary.

%%%%%%%%%%%%%%%%%%%%%%%%%%%%%%%%%%%%%%%%%%%%%%%%%%%%%%%%%%%%%%%%%%%%%%%%%%%%
%%%%%%%%%%%%%%%%%%%%%%%%%%%%%%%%%%%%%%%%%%%%%%%%%%%%%%%%%%%%%%%%%%%%%%%%%%%%
\section{Quantum brachistochrone}
\label{sec:qb}

%%%%%%%%%%%%%%%%%%%%%%%%%%%%%%%%%%%%%%%%%%%%%%%%%%%%%%%%%%%%%%%%%%%%%%%%%%%%
\subsection{Quantum brachistochrone equation}

In order to derive the optimal solution of the time-dependent 
Hamiltonian $H(t)$ and state $|\psi(t)\rangle$, 
we define the action
\be
 S = \int_0^T dt\,\left(L_{\rm T}+L_{\rm S}+L_{\rm C}\right),
\ee
as an integral between the initial and 
final time~\cite{CHKO1} .
Each term is expressed as 
\be
 & & L_{\rm T} = \frac{\sqrt{\la\dot{\psi}|(1-P)|\dot{\psi}\ra}}{\Delta E}, \\
 & & L_{\rm S} = i\la\phi|\dot{\psi}\ra
 -\la\phi|H|\psi\ra
 +({\rm h.c.}), \\
 & & L_{\rm C} = \sum_a \lambda_a f_a(H).
\ee
We use notations $P = |\psi\ra\la\psi|$, 
$\Delta E^2 = \la\psi|H^2|\psi\ra - \la\psi|H|\psi\ra^2$ 
and dotted symbol for the time derivative.
$L_{\rm T}dt$ represents the Fubini--Study distance 
$\langle d\psi|(1-P)|d\psi\rangle$
divided by $\Delta E$. 
The energy variance $\Delta E$ plays the role of the velocity 
in quantum systems~\cite{AA}, 
which means that $\int L_{\rm T}dt$ represents the time to be taken 
for the evolution of the quantum state.
The other two terms are introduced to impose constraints 
which must hold throughout the evolution.
The Schr\"odinger equation $i|\dot{\psi}\ra=H|\psi\ra$ 
is implemented by $L_{\rm S}$ and  
other constraints $f_a(H)=0$ by $L_{\rm C}$.
$|\phi\ra$ and $\lambda_a$ represent the Lagrange multipliers.
They are determined in course of the calculation.

For fixed initial state $|\psi(0)\ra$ and final one $|\psi(T)\ra$,
the action is extremized by variations 
$H\to H+\delta H$ and $|\psi\ra\to|\psi\ra+|\delta\psi\ra$.
Setting the linear term to be zero, we obtain 
\be
 & & 
 i\frac{d}{dt}\left(\frac{H-\la H\ra}{2\Delta E^2}\right)|\psi\ra
 +\left(i\frac{d}{dt}-H\right)|\phi\ra =0, \\
 & &  
 -\frac{1}{2\Delta E^2}\left(HP+PH-2P\la H\ra\right)
 -\left(|\psi\ra\la\phi|+|\phi\ra\la\psi|\right)
 +F =0, 
\ee
where 
\be
 F = \sum_a \lambda_a f_a'(H).
\ee
The prime denotes the derivative with respect to $H$.
After some calculations, we obtain 
\be
 F(t)= U(t)\left[|\tilde{\phi}(0)\rangle\langle\psi(0)|
 +|\psi(0)\rangle\langle\tilde{\phi}(0)|\right]U^\dag(t),
\ee
where $U(t)$ is the time-evolution operator 
for the Hamiltonian $H(t)$ and 
\be
 |\tilde{\phi}(0)\rangle=|\phi(0)\rangle
 +\frac{H(0)-\langle\psi(0)|H(0)|\psi(0)\rangle}{2\Delta E^2(0)}|\psi(0)\rangle.
\ee
Thus, $F$ satisfies the QB equation~\cite{CHKO1}
\be
 & & i\dot{F}(t)=[H(t),F(t)], \label{qb1}
\ee
with initial condition
\be
 & & \left(1-P(0)\right)F(0)\left(1-P(0)\right)=0. \label{qb2}
\ee
We note that the condition (\ref{qb2}) is unchanged 
throughout the evolution.
These equations are solved under fixed $T$.
Then, we minimize the passage time $T$ 
to accomplish our purpose to obtain the optimal solution.

For convenience of the calculation, 
we rewrite the QB equation
by using the basis operators $\{X_\mu\}_{\mu=0,1,\cdots,N^2-1}$
where $N$ is the dimension of the Hilbert space.
These satisfy the orthonormal relation
\be
 \frac{1}{N}\Tr X_\mu X_\nu = \delta_{\mu\nu},
\ee
and the completeness relation
\be
 \frac{1}{N}\sum_{\mu=0}^{N^2-1}(X_\mu)_{ij}(X_\mu)_{kl}=\delta_{il}\delta_{jk}.
\ee
The structure constant $f_{\mu\nu\rho}$ is defined 
by the commutation relation 
\be
 [X_\mu,X_\nu]=i\sum_\rho f_{\mu\nu\rho}X_\rho.
\ee
We set $X_0$ to be the identity operator and 
use the vector representation $\bm{X}$ 
for other components $\{X_a\}_{a=1,\cdots,N^2-1}$.
Then, the Hamiltonian is represented as 
\be
 H(t)=h_0(t)+\bm{h}(t)\cdot\bm{X}. \label{HX}
\ee
We note that the first term does not bring any quantum effect 
and can be eliminated by a proper gauge transformation
$|\psi(t)\ra \to e^{i\varphi(t)}|\psi(t)\ra$.

Using this representation, we can parametrize $P$ and $F$ as 
\be
 & & P(t)=\frac{1}{N}+\frac{\sqrt{N-1}}{N}\bm{e}(t)\cdot\bm{X}, \\
 & & F(t)= \lambda\left(
 \frac{\bm{l}(0)\cdot\bm{e}(0)}{\sqrt{N-1}}
 +\bm{l}(t)\cdot\bm{X}\right), \label{FX}
\ee
where $\bm{e}$ and $\bm{l}$ are normalized vectors 
with the angle between them fixed:
$\bm{l}(t)\cdot\bm{e}(t)=\bm{l}(0)\cdot\bm{e}(0)$.
These representations are derived from 
$\Tr P=\Tr P^2=1$ and $\Tr F=\Tr FP$.
In the following analysis, 
the overall constant $\lambda$ is not important and 
is left undetermined.
We also note that the first term in (\ref{FX}) is derived from 
the condition that the first term in (\ref{HX}) is fixed and 
is not important as well. 
From equation (\ref{qb1}), we can write 
the equations of motion as 
\be
 & & \dot{\bm{l}}(t) = \bm{h}(t)\times\bm{l}(t), \label{lambdaeq}
\ee
where the vector product is defined by 
$(\bm{h}\times\bm{l})_a=\sum_{b,c}f_{abc}h_bl_c$.
Since the state $|\psi(t)\rangle$ satisfies 
the Schr\"odinger equation, $P$ follows the same equation as (\ref{qb1})
and, as a result,  $\dot{\bm{e}}(t)=\bm{h}(t)\times \bm{e}(t)$.

The solution of the QB equation strongly depends on constraints to impose.
We consider the case where components $a=1,2,\cdots, k$ are fixed as 
\be
 H_{\rm C}(t)= \sum_{a=1}^k h_a^{(0)}(t)X_a. \label{hconstraint}
\ee
Correspondingly, $f_a$ and $F$ are given by 
\be
 & & f_a = \Tr H(t)X_a -Nh_a^{(0)}(t) \qquad (a=1,2,\cdots,k), \\
 & & F(t) = \lambda \sum_{a=1}^k l_a(t)X_a.
\ee
The QB equation is solved to give other components: 
\be
 & & H_{\rm QB}(t)= \sum_{p=k+1}^{N^2-1}h_p^{(1)}(t)X_p.
\ee
The total Hamiltonian is given by $H(t)=H_{\rm C}(t)+H_{\rm QB}(t)$.
For a given $h_a^{(0)}(t)$, 
time-dependence of $l_a(t)$ and $h^{(1)}_p(t)$ are obtained by 
solving equation (\ref{lambdaeq})
and the initial condition is determined by equation (\ref{qb2}).

%%%%%%%%%%%%%%%%%%%%%%%%%%%%%%%%%%%%%%%%%%%%%%%%%%%%%%%%%%%%%%%%%%%%%%%%%%%%
\subsection{Lewis-Riesenfeld invariant and transitionless quantum driving}

The QB equation is written in terms of 
the operator $F=\sum_a\lambda_af_a'(H)$ 
coming from the constraints $f_a(H)=0$.
It is crucial in the following analysis 
to notice that this quantity $F$ is nothing but 
the Lewis--Riesenfeld invariant~\cite{LR}.
It satisfies equation (\ref{qb1}) and, 
as a result, all eigenvalues of $F$ become independent of time.
We can write 
\be
 F(t) = \sum_{n}\lambda_n |n(t)\ra\la n(t)|,
\ee
where the time-independent eigenvalue is denoted by $\lambda_n$ and 
$|n(t)\ra$ is the eigenstate defined at each time.
By using the eigenstates, we can write the state as 
\be
 |\psi(t)\ra = \sum_n c_n e^{i\alpha_n(t)}|n(t)\ra, \label{ad}
\ee
with a phase factor
\be
 \alpha_n(t)=\int_0^t dt'\,\langle n(t')|
 \left(i\frac{d}{dt'}-H(t')\right)
 |n(t')\rangle.
\ee
The real constant $c_n$ is shown to be independent of time, 
which means that if we start the evolution from $|n(0)\ra$, 
the state remains the eigenstate $|n(t)\ra$ at arbitrary $t$.

In order to represent the optimal Hamiltonian 
in the basis of $\{|n(t)\ra\}$,
we multiply $\la m(t)|$ from the left and 
$|n(t)\ra$ from the right to equation (\ref{qb1}).
We assume that there is no degeneracy in the spectrum
for simplicity.
Then, the off-diagonal element $\langle m(t)|H(t)|n(t)\rangle$
with $m\ne n$ is calculated as
\be
 \langle m(t)|H(t)|n(t)\rangle
 = i\langle m(t)|\dot{n}(t)\rangle.
\ee
The Hamiltonian is thus written as 
$H(t)=H_0(t)+H_1(t)$ where
\be
 & & H_0(t) =  \sum_{n}E_n(t)|n(t)\ra\la n(t)|, \label{tqd1} \\
 & & H_1(t) = i\sum_{m\ne n}
 |m(t)\ra\la m(t)|\dot{n}(t)\ra\la n(t)|. \label{tqd2}
\ee
The diagonal part $H_0(t)$ commutes with the invariant $F(t)$.
Its eigenvalue $E_n(t)$ cannot be specified from equation (\ref{qb1}).
We see that this Hamiltonian represents the formula of 
the transitionless quantum driving~\cite{Berry}.
In that method, 
the adiabatic state (\ref{ad}), with $H(t)$ replaced by $H_0(t)$,  
is constructed for a given Hamiltonian $H_0(t)$.
This is not the solution of the Schr\"odinger equation and 
we apply the counter-diabatic Hamiltonian $H_1(t)$
to make the system transitionless. 
Then, the adiabatic state becomes the exact solution of the equation.

The result here shows that 
the assisted AP, transitionless quantum driving and 
shortcuts to adiabaticity are essentially equivalent and
they can be derived from the QB equation.

%%%%%%%%%%%%%%%%%%%%%%%%%%%%%%%%%%%%%%%%%%%%%%%%%%%%%%%%%%%%%%%%%%%%%%%%%%%%
\subsection{Adiabatic passage as a quantum brachistochrone trajectory}

Although we have shown that the solution of the QB equation 
can be interpreted as the AP, 
it is not clear from the beginning what kind of adiabatic state is obtained.
It will be practically useful if the adiabatic state is determined 
by the constrained part $H_{\rm C}(t)$
so that we can control the system at will.
That is, we want to know the cases where 
the relations $H_0(t)=H_{\rm C}$ and $H_1(t)=H_{\rm QB}$ hold.
In the following, we set the initial state $|\psi(0)\rangle$
to be an eigenstate of $F(0)$.
In this case, we have 
$|\phi(t)\rangle \propto |\psi(t)\rangle$ and 
$F(t)\propto P(t)$.

The condition that $H_{\rm C}(t)$ determines the adiabatic state 
is given by the commutation relation
\be
 [H_{\rm C}(t),F(t)]=0.  \label{hcf}
\ee
This can be written as 
\be
 \bm{h}^{(0)}(t)\times\bm{l}(t) = 0.
\ee
We easily see that a possible solution is given 
by $\bm{l}(t)\propto\bm{h}^{(0)}(t)$.
It is hard to imagine that the equation has other nontrivial solutions
since the commutation relation (\ref{hcf}) must hold for arbitrary $t$.
Although it is an interesting problem to find such a solution
in a specific example,
we discuss the solution $\bm{l}(t)\propto\bm{h}^{(0)}(t)$
in the following general analysis.
Thus, $F$ is equivalent to $H_{\rm C}(t)$ as
\be
 F(t)=\frac{1}{h^{(0)}(t)}\sum_{a=1}^k h_a^{(0)}(t)X_a
 =\frac{1}{h^{(0)}(t)}H_{\rm C}(t), \label{fhc}
\ee
where $h^{(0)}(t)=|\bm{h}^{(0)}(t)|=\sqrt{\sum_ah_a^{(0)}(t)h_a^{(0)}(t)}$.
We neglected the unimportant constant term and overall factor.
The QB equation is given by 
\be
 & & \dot{\bm{h}}^{(0)}(t)-\frac{\dot{h}^{(0)}(t)}{h^{(0)}(t)}\bm{h}^{(0)}(t)
 =\bm{h}^{(1)}(t)\times\bm{h}^{(0)}(t).
 \label{qbtqd}
\ee
We note that this equation does not necessarily have the solution.
If no solution exists, $H_{0}(t)\ne H_{\rm C}(t)$ is implied.
When the solution exists, each part of the Hamiltonian is given by
\be
 & & H_{\rm C}(t)=H_0(t), \\
 & & H_{\rm QB}(t)=H_1(t)+\delta H_1(t), \label{deltaH1}
\ee
 where
\be
 [H_0(t),\delta H_1(t)]=0.
\ee
Generally, there exists the Hamiltonian $\delta H_1(t)$ in $H_{\rm QB}(t)$
such that the adiabatic state is not disturbed.
Since the QB equation is the most general equation, 
the result depends strongly on constraints to impose.
If the constraints are too loose, the solution has many ambiguities 
and is not determined uniquely.
On the other hand, tight constraints will give 
trivial and ineffective solutions.
It is important to choose proper constraints 
to obtain the nontrivial results.
Since it is hard to study general conditions, 
we examine the QB equation explicitly 
in the next section.

%%%%%%%%%%%%%%%%%%%%%%%%%%%%%%%%%%%%%%%%%%%%%%%%%%%%%%%%%%%%%%%%%%%%%%%%%%%%
%%%%%%%%%%%%%%%%%%%%%%%%%%%%%%%%%%%%%%%%%%%%%%%%%%%%%%%%%%%%%%%%%%%%%%%%%%%%
\section{Examples}
\label{sec:ex}

%%%%%%%%%%%%%%%%%%%%%%%%%%%%%%%%%%%%%%%%%%%%%%%%%%%%%%%%%%%%%%%%%%%%%%%%%%%%
\subsection{$N=2$}

First, we consider the simplest example of $N=2$.
The Hamiltonian is written by the Pauli matrices $\bm{\sigma}$ as 
\be
 H(t) = \frac{1}{2}\bm{h}(t)\cdot\bm{\sigma}
 = \frac{1}{2}\bmat{cc} h_3(t) & h_1(t)-ih_2(t) \\ 
 h_1(t)+ih_2(t) & -h_3(t) \emat.
\ee
The basis operator is given by 
$\bm{X}=(1,\sigma_1,\sigma_2,\sigma_3)$ and 
the structure constant by $f_{abc}=2\epsilon_{abc}$.
Equation (\ref{qbtqd}) is rewritten as  
\be
  & & \dot{\bm{h}}^{(0)}(t)-\frac{\dot{h}^{(0)}(t)}{h^{(0)}(t)}\bm{h}^{(0)}(t)
 =\bm{h}^{(1)}(t)\times\bm{h}^{(0)}(t), 
\ee
where the definition of the vector product is changed as
$(\bm{h}^{(1)}\times\bm{h}^{(0)})_a=\sum_{bc}\epsilon_{abc}h_b^{(1)}h_c^{(0)}$.
Considering the vector product with $\bm{h}^{(0)}(t)$, we obtain
\be
 \bm{h}^{(1)}(t)=
 \frac{\bm{h}^{(0)}(t)\times\dot{\bm{h}}^{(0)}(t)}{|\bm{h}^{(0)}(t)|^2}.
 \label{h1}
\ee
Here, we used the property of the antisymmetric tensor 
$\sum_a\epsilon_{abc}\epsilon_{ade}=\delta_{bd}\delta_{ce}-\delta_{be}\delta_{cd}$.
This result gives the counter-diabatic part $H_1(t)$ derived 
in the method of the transitionless quantum driving~\cite{Berry}.
That is, we conclude that 
$H_{\rm C}(t)=H_0(t)$ and $H_{\rm QB}(t)=H_1(t)$ in the present case.

We note that the nontrivial result is obtained only in the case 
where two components of $\bm{h}^{(0)}$ are constrained.
In that case, the third component is determined by equation (\ref{h1}).
When one component of $\bm{h}^{(0)}$ is constrained,
no quantum effect appears and we do not have any interesting result.
For three components fixed, the Hamiltonian is completely specified 
and there is no room for finding the optimal Hamiltonian. 

%%%%%%%%%%%%%%%%%%%%%%%%%%%%%%%%%%%%%%%%%%%%%%%%%%%%%%%%%%%%%%%%%%%%%%%%%%%%%%%%%%
\subsection{$N=3$}

Next, we consider the case of $N=3$.
In this case, we use the Gell-Mann matrices 
\be
 & & \lambda_1 = \bmat{ccc} 0 & 1 & 0 \\ 1 & 0 & 0 \\ 0 & 0 & 0 \emat, \quad
 \lambda_2 = \bmat{ccc} 0 & -i & 0 \\ i & 0 & 0 \\ 0 & 0 & 0 \emat, \quad
 \lambda_3 = \bmat{ccc} 1 & 0 & 0 \\ 0 & -1 & 0 \\ 0 & 0 & 0 \emat, \no\\
 & & \lambda_4 = \bmat{ccc} 0 & 0 & 1 \\ 0 & 0 & 0 \\ 1 & 0 & 0 \emat, \quad
 \lambda_5 = \bmat{ccc} 0 & 0 & -i \\ 0 & 0 & 0 \\ i & 0 & 0 \emat, \quad
 \lambda_6 = \bmat{ccc} 0 & 0 & 0 \\ 0 & 0 & 1 \\ 0 & 1 & 0 \emat, \no\\
 & & \lambda_7 = \bmat{ccc} 0 & 0 & 0 \\ 0 & 0 & -i \\ 0 & i & 0 \emat, \quad
 \lambda_8 = \frac{1}{\sqrt{3}}
 \bmat{ccc} 1 & 0 & 0 \\ 0 & 1 & 0 \\ 0 & 0 & -2 \emat,
 \label{gellmann}
\ee
as the basis operators.
Normalizing the operators, we have $X_{a}=\frac{\sqrt{6}}{2}\lambda_a$ and
the structure constant is given in table~\ref{table:sc}.
Since we do not have useful relations on the structure constant, 
it is hard to consider the general properties
as we did in the case of $N=2$.
In the following, we study properties of the solution explicitly.
%%%%%%%%%%%%%%%%%%
\begin{table}[t]
\begin{center}
\begin{tabular}{|ccc|c||ccc|c|}
\hline
 $a$ & $b$ & $c$ & $f_{abc}$ & $a$ & $b$ & $c$ & $f_{abc}$ \\
\hline
 1 & 2 & 3 & $\sqrt{6}$  & 3 & 4 & 5 & $\frac{1}{2}\sqrt{6}$ \\
 1 & 4 & 7 & $\frac{1}{2}\sqrt{6}$  & 3 & 6 & 7 & -$\frac{1}{2}\sqrt{6}$ \\
 1 & 5 & 6 & -$\frac{1}{2}\sqrt{6}$ & 4 & 5 & 8 & $\frac{\sqrt{3}}{2}\sqrt{6}$ \\
 2 & 4 & 6 & $\frac{1}{2}\sqrt{6}$  & 6 & 7 & 8 & $\frac{\sqrt{3}}{2}\sqrt{6}$ \\
 2 & 5 & 7 & $\frac{1}{2}\sqrt{6}$  & & & & \\
\hline
\end{tabular}
\caption{The structure constant $f_{abc}$ for 
the basis operators $X_a=\frac{\sqrt{6}}{2}\lambda_a$ ($a=1,\cdots 8$)
at $N=3$.
$\lambda_a$ represent the Gell-Mann matrices in equation~(\ref{gellmann}).
$f_{abc}$ is antisymmetric with respect to all pairs of indices.
The other components not shown in the table are zero.
}
\label{table:sc}
\end{center}
\end{table}
%%%%%%%%%%%%%%%%%%

%%%%%%%%%%%%%%%%%%%%%%%%%%%%%%%%%%%%%%%%%%%%%%%%%%%%%%%%%%%%%%%%%%%%%%%%%%%%%%%%%%
\subsubsection{$k=2$.}

We first consider the case where two components of $\bm{h}(t)$, 
$a$ and $b$, are fixed.
We assume that there exists components $p$ such that $f_{abp}\ne 0$.
Otherwise, no quantum effect appears.
The explicit analysis depends on how many of $p$ exist.
Although it is always possible to choose basis operators 
such that the unique $p$ exists, 
we consider the case when the Gell-Mann matrices are used as 
the basis operators.
In that case, we have the following two patterns.

%%%%%%%%%%%%%%%%%%
\begin{itemize}
\item{$a=1$, $b=2$.}

In this case, $p=3$ is the only component with $f_{abp}\ne 0$.
From equation (\ref{qbtqd}), we obtain
\be
 f_{123} h_3^{(1)}(t)
 =\frac{h_1^{(0)}(t)\dot{h}_2^{(0)}(t)-h_2^{(0)}(t)\dot{h}_1^{(0)}(t)}
 {|\bm{h}^{(0)}|^2}.
\ee
We also see that
$h_8^{(1)}(t)$ is left undetermined from the QB equation.
In fact, $X_8$ commutes with $H_{\rm C}(t)=h_1^{(0)}X_1+h_2^{(0)}X_2$ 
and does not disturb the adiabatic state.
On the other hand, 
by using the formula of the transitionless quantum driving, 
equations (\ref{tqd1}) and (\ref{tqd2}), 
we obtain
the counter-diabatic Hamiltonian $H_1(t)=h_3^{(1)}(t)X_3$.
We conclude that the AP is realized as 
\be
 & & H_{\rm C}(t)=H_{0}(t)=h_1^{(0)}(t)X_1+h_2^{(0)}(t)X_2, \\
 & & H_{\rm QB}(t)=H_1(t)+h_8^{(1)}(t)X_8=h_3^{(1)}(t)X_3+h_8^{(1)}(t)X_8.
\ee

%%%%%%%%%%%%%%%%%%%
\item{$a=4$, $b=5$.}

There exists two components $f_{345}$ and $f_{458}$ with nonzero value.
Then, we obtain 
\be
 f_{345} h_3^{(1)}(t)+f_{458} h_8^{(1)}(t)
 =\frac{h_4^{(0)}(t)\dot{h}_5^{(0)}(t)-h_4^{(0)}(t)\dot{h}_5^{(0)}(t)}
 {|\bm{h}^{(0)}|^2}.
\ee
The solution is given by 
\be
 & & h_3^{(1)}(t)=\frac{1}{2\sqrt{6}}
 \frac{h_4^{(0)}(t)\dot{h}_5^{(0)}(t)-h_4^{(0)}(t)\dot{h}_5^{(0)}(t)}
 {|\bm{h}^{(0)}|^2}
 +\sqrt{3}h(t),
 \\
 & & h_8^{(1)}(t)=\frac{\sqrt{3}}{2\sqrt{6}}
 \frac{h_4^{(0)}(t)\dot{h}_5^{(0)}(t)-h_4^{(0)}(t)\dot{h}_5^{(0)}(t)}
 {|\bm{h}^{(0)}|^2}
 -h(t),
\ee
where $h(t)$ is an arbitrary function.
$h(t)(\sqrt{3}X_3-X_8)$ commutes with the constraint part
and does not affect the adiabatic part.
Each first term in the above equations is derived from 
equations (\ref{tqd1}) and (\ref{tqd2}).
Therefore, we conclude also in this case as 
\be
 & & H_{\rm C}(t)=h_4^{(0)}(t)X_4+h_5^{(0)}(t)X_5, \\
 & & H_{\rm QB}(t)
 =h_3^{(1)}(t)X_3+h_8^{(1)}(t)X_8+h(t)(\sqrt{3}X_3-X_8).
\ee

\end{itemize}

%%%%%%%%%%%%%%%%%%%%%%%%%%%%%%%%%%%%%%%%%%%%%%%%%%%%%%%%%%%%%%%%%%%%%%%%%%%%%%%%%%
\subsubsection{$k=3$.}

We next consider the case where three components $a$, $b$ and $c$ are fixed.
In this case, possible numbers of components $p$ 
such that $f_{abp}\ne 0$ are four at the maximum.
We can consider each pattern 
to find the formula of the transitionless driving.
In the present case, another problem arises such that  
the structure constant among the constraint components is nonzero: 
$f_{abc}\ne 0$.

%%%%%%%%%%%%%%%%%%%%%%%%%
\begin{itemize}
\item{$a=1$, $b=2$, $c=4$.}

In this case, 
components 3, 6, 7 and 8 participate to equations 
to determine $\bm{h}^{(1)}$.
The equations to be solved are given by 
\be
 & & -f_{123}h_2^{(0)}h_3^{(1)}-f_{147}h_4^{(0)}h_7^{(1)}
 =\dot{h}_1^{(0)}-\frac{\dot{h}^{(0)}}{h^{(0)}}h_1^{(0)}, \\
 & & f_{123}h_1^{(0)}h_3^{(1)}-f_{246}h_4^{(0)}h_6^{(1)} 
 =\dot{h}_2^{(0)}-\frac{\dot{h}^{(0)}}{h^{(0)}}h_2^{(0)}, \\
 & & f_{246}h_2^{(0)}h_6^{(1)}+f_{147}h_1^{(0)}h_7^{(1)}
 =\dot{h}_4^{(0)}-\frac{\dot{h}^{(0)}}{h^{(0)}}h_4^{(0)}, \\
 & & f_{345}h_4^{(0)}h_3^{(1)}+f_{156}h_1^{(0)}h_6^{(1)}
 +f_{257}h_2^{(0)}h_7^{(1)}+f_{458}h_4^{(0)}h_8^{(1)} =  0.
\ee
The third equation is derived from the first and second equations.
Since three independent equations are imposed on four variables 
$h_3^{(1)}$, $h_6^{(1)}$, $h_7^{(1)}$ and $h_8^{(1)}$, 
we have an arbitrariness in choosing the solution 
as in the previous example.

On the other hand, if we apply the formula of 
the transitionless quantum driving
by setting $H_0(t)=H_{\rm C}(t)=h_1^{(0)}(t)X_1+h_2^{(0)}(t)X_2+h_4^{(0)}(t)X_4$, 
we obtain the counter-diabatic Hamiltonian 
\be
 & & H_1(t) = h_3^{(1)}(t)X_3+h_6^{(1)}(t)X_6+h_7^{(1)}(t)X_7+h_8^{(1)}(t)X_8, \\
 & & h_3^{(1)}= \frac{1}{\sqrt{6}}
 \left(1+\frac{3h_4^{(0)2}}{2|\bm{h}^{(0)}|^2}\right)
 \frac{h_1^{(0)}\dot{h}_2^{(0)}-h_2^{(0)}\dot{h}_1^{(0)}}
 {|\bm{h}^{(0)}|^2}, \\
 & & h_6^{(1)}= \sqrt{\frac{3}{2}}\frac{h_4^{(0)}h_1^{(0)}}{|\bm{h}^{(0)}|^2}
 \frac{h_1^{(0)}\dot{h}_2^{(0)}-h_2^{(0)}\dot{h}_1^{(0)}}
 {|\bm{h}^{(0)}|^2}
 +\sqrt{\frac{2}{3}}
 \frac{h_2^{(0)}\dot{h}_4^{(0)}-h_4^{(0)}\dot{h}_2^{(0)}}
 {|\bm{h}^{(0)}|^2}, \\
 & & h_7^{(1)}= -\sqrt{\frac{3}{2}}\frac{h_4^{(0)}h_2^{(0)}}{|\bm{h}^{(0)}|^2}
 \frac{h_1^{(0)}\dot{h}_2^{(0)}-h_2^{(0)}\dot{h}_1^{(0)}}
 {|\bm{h}^{(0)}|^2}
 -\sqrt{\frac{2}{3}}
 \frac{h_4^{(0)}\dot{h}_1^{(0)}-h_1^{(0)}\dot{h}_4^{(0)}}{|\bm{h}^{(0)}|^2}, \\
 & & h_8^{(1)}= -\frac{3\sqrt{2}}{4}
 \frac{h_4^{(0)2}}{|\bm{h}^{(0)}|^2}
 \frac{h_1^{(0)}\dot{h}_2^{(0)}-h_2^{(0)}\dot{h}_1^{(0)}}{|\bm{h}^{(0)}|^2}.
\ee
This result is a solution of the QB equation.
The arbitrariness comes from the choice of the diagonal part of $H_1$,
$\delta H_1(t)$ in equation (\ref{deltaH1}).

%%%%%%%%%%%%%%%%%%%%%%%%%%
\item{$a=3$, $b=4$, $c=5$.}

This case is different from the previous one
due to the property $f_{345}\ne 0$.
Then, the ansatz $\bm{l}(t)\propto\bm{h}^{(0)}(t)$ does not solve 
the QB equation.
Equation (\ref{lambdaeq}) is written explicitly as
\be
 & & \dot{l}_3 = f_{345}\left(h_4^{(0)}l_5-h_5^{(0)}l_4\right), \\
 & & \dot{l}_4 = f_{345}\left(h_5^{(0)}l_3-h_3^{(0)}l_5\right)-f_{458}h_8^{(1)}l_5, \\
 & & \dot{l}_5 = f_{345}\left(h_3^{(0)}l_4-h_4^{(0)}l_3\right)+f_{458}h_8^{(1)}l_4, \\
 & & 0 =  f_{123}h_2^{(1)}l_3-f_{147}h_7^{(1)}l_4-f_{156}h_6^{(1)}l_5, \\
 & & 0 =  -f_{123}h_1^{(1)}l_3-f_{246}h_6^{(1)}l_4-f_{257}h_7^{(1)}l_5, \\
 & & 0 =  f_{246}h_2^{(1)}l_4-f_{367}h_7^{(1)}l_3+f_{156}h_1^{(1)}l_5, \\
 & & 0 =  f_{147}h_1^{(1)}l_4+f_{257}h_2^{(1)}l_5-f_{367}h_6^{(1)}l_3, \\
 & & 0 = f_{458}\left(h_4^{(0)}l_5-h_5^{(0)}l_4\right).
\ee
From the first and last equations, we obtain 
\be
 \left(l_3,l_4,l_5\right)
 = \left(
 \sin\theta, \frac{h_4^{(0)}}{\sqrt{h_4^{(0)2}+h_5^{(0)2}}}\cos\theta, 
 \frac{h_5^{(0)}}{\sqrt{h_4^{(0)2}+h_5^{(0)2}}}\cos\theta
 \right),
\ee
where $\theta$ is a constant.
Using the second or third equations, we can calculate $h_8^{(1)}$ as
\be
 h_8^{(1)}=\frac{f_{345}}{f_{458}}\left(
 \frac{1}{f_{345}}\frac{h_4^{(0)}\dot{h}_5^{(0)}
 -h_5^{(0)}\dot{h}_4^{(0)}}{h_4^{(0)2}+h_5^{(0)2}}
 -h_3^{(0)} +\sqrt{h_4^{(0)2}+h_5^{(0)2}}\tan\theta
 \right). \no\\
\ee
The other components $h_1^{(1)}$, $h_2^{(1)}$, $h_6^{(1)}$ and $h_7^{(1)}$
are shown to be zero.
Thus, we obtain  $H_{\rm QB}(t) = h_8^{(1)}(t)X_8$.

Correspondingly, for $H_0(t)=h^{(0)}_3X_3+h^{(0)}_4X_4+h^{(0)}_5X_5$, 
the AP is given by the counter-diabatic Hamiltonian of the form 
$H_1(t)=h_3^{(1)}(t)X_3+h_4^{(1)}(t)X_4+h_5^{(1)}(t)X_5+h_8^{(1)}(t)X_8$.
Due to the property $f_{345}\ne 0$, $H_1$ inevitably 
has the same components as $H_0$.
Therefore, in this case, $H_0(t)\ne H_{\rm C}(t)$.

\end{itemize}

In the same way, we can show that 
the formula of the transitionless driving is derived when $f_{abc}=0$.
We have examined four possible patterns depending on the number of 
components $p$ such that $f_{abp}\ne 0$.
In some cases, the QB equation cannot determine the solution uniquely 
and there exists some ambiguity.
It does not disturb the adiabatic state and
we can conclude that the constraint part gives the adiabatic state.
This is generalized to the case where higher numbers of 
components are constrained.

%%%%%%%%%%%%%%%%%%%%%%%%%%%%%%%%%%%%%%%%%%%%%%%%%%%%%%%%%%%%%%%%%%%%%%%%%%%%
%%%%%%%%%%%%%%%%%%%%%%%%%%%%%%%%%%%%%%%%%%%%%%%%%%%%%%%%%%%%%%%%%%%%%%%%%%%%
\section{Stability of the adiabatic passage}
\label{sec:stab}

%%%%%%%%%%%%%%%%%%%%%%%%%%%%%%%%%%%%%%%%%%%%%%%%%%%%%%%%%%%%%%%%%%%%%%%%%%%%
\subsection{General consideration}

The advantage of describing the AP as the QB 
is that one can study the stability of the driving.
It is shown in several examples that the AP is robust against
parameter variations~\cite{Betal, KT}.
Here, we show that it is generally correct. 
We also find that the instability arises 
when we consider perturbation by different kinds of operators.

The QB equation is derived by expanding the action 
up to first order in $\delta H$ and $|\delta\psi\ra$.
The stability of the extremized solution is found 
by examining the second order. 
As a possible situation, 
we consider the case when the Hamiltonian is changed 
as $H\to H+\delta H$.
$H$ and $|\psi\rangle$ are determined from the QB equation and 
we see what happens if the counter-diabatic Hamiltonian deviates from 
the ideal form.

The second order of the action in $\delta H$ is calculated as 
\be
 I(t) &=& -\frac{1}{2\Delta E^2(t)}\left(
 \la\delta H^2(t)\ra- \la\delta H(t)\ra^2\right) \no\\
 && +\frac{3}{8}\frac{1}{\Delta E^4(t)}\Bigl(
 \la H(t)\delta H(t)+\delta H(t) H(t)\ra 
 -2\la\delta H(t)\ra\la H(t)\ra\Bigr)^2,
 \label{stab1}
\ee
where $\la(\cdots)\ra=\la\psi(t)|(\cdots)|\psi(t)\ra$.
The solution of the QB equation is stable 
when $I(t)> 0$.
This is the general result applied to any QB trajectories.
We are interested in the case where $H(t)=H_0(t)+H_1(t)$ with 
$H_0(t)$ is the Hamiltonian giving the adiabatic state
and $H_1(t)$ is the counter-diabatic Hamiltonian.
In this case, the state $|\psi\rangle$ is the eigenstate of $H_0(t)$
and the diagonal elements of $H_1(t)$ are zero in this representation.
Then, we can write 
\be
 I(t) = -\frac{\la\delta H^2(t)\ra- \la\delta H(t)\ra^2}{2\la H_1^2(t)\ra}
 +\frac{3}{8}
 \frac{\la H_1(t)\delta H(t)+\delta H(t) H_1(t)\ra^2}{\la H_1^2(t)\ra^2}.
 \label{stab2}
\ee
This is the main result in this section.
The stability of the AP is given by the condition $I(t)>0$.

When the variation is proportional to the counter-diabatic Hamiltonian as 
$\delta H(t)=c(t) H_1(t)$ with an arbitrary scalar function $c(t)$,
we can show that $I(t) = c^2(t)$, which means that the AP is stable against 
any parameter variations in the counter-diabatic Hamiltonian.
This result is consistent with previous examples 
showing stable driving~\cite{Betal, KT}.
It is also possible to consider the unstable perturbation in principle.
Such a perturbation is realized when the second term of equation (\ref{stab2}) 
becomes smaller compared with the first term.
It is accomplished when 
$H_1(t)$ and $\delta H(t)$ anticommute with each other: 
$H_1(t)\delta H(t)+\delta H(t) H_1(t) = 0$.
We can generally say that the AP is unstable against 
perturbations by operators anticommuting with the counter-diabatic Hamiltonian.

%%%%%%%%%%%%%%%%%%%%%%%%%%%%%%%%%%%%%%%%%%%%%%%%%%%%%%%%%%%%%%%%%%%%%%%%%%%%
\subsection{Example: $N=3$}

Equation $I(t)> 0$ represents a necessary condition 
of the stability.
We examine this condition by using an example
to see whether the expected behavior is obtained.
Although the simplest example of the quantum system is 
the two-level system $N=2$, 
unstable perturbations cannot be considered in this case.
If we fix two components of the magnetic field, 
the third component is determined by the QB equation and 
no other components exist.
For this reason, we treat the three-level system $N=3$ in this subsection.

As an adiabatic Hamiltonian, we consider the magnetic field in the $xy$ plane
\be
 & & H_0(t) = h_1(t)S_1+h_2(t)S_2, \\
 & & h_1(t)=h_0\cos\omega t, \\
 & & h_2(t)=h_0\sin\omega t,
\ee
where $S_1$ and $S_2$ are spin operators
\be
 & & S_1 = \frac{1}{\sqrt{2}}
 \bmat{ccc} 0 & 1 & 0 \\ 1 & 0 & 1 \\ 0 & 1 & 0 \emat, \\
 & & S_2 = \frac{1}{\sqrt{2}}
 \bmat{ccc} 0 & -i & 0 \\ i & 0 & -i \\ 0 & i & 0 \emat, 
\ee
and can be expressed by 
linear combinations of the Gell-Mann matrices.
The adiabatic state is given by 
\be
 |\psi_{\rm ad}(t)\ra = \frac{1}{2}
 \bmat{cc} e^{-i\omega t} \\ \sqrt{2} \\ e^{i\omega t} \emat.
 \label{ad2}
\ee
By using the formulas (\ref{tqd1}) and (\ref{tqd2})
of the transitionless quantum driving, 
we obtain the counter-diabatic Hamiltonian 
\be
 H_1(t)=\omega S_3 
 =\omega\bmat{ccc} 1 & 0 & 0 \\ 0 & 0 & 0 \\ 0 & 0 & -1\emat.
\ee
From the general consideration, the driving should be stable 
against the variation $\omega\to\omega+\delta\omega(t)$ where $\delta\omega(t)$
is an arbitrary function.
It is also possible to consider perturbations inducing instabilities.
For example, as operators which anticommute with $S_3$,
we can use the Gell-Mann matrices $\lambda_4$ and $\lambda_5$.

We numerically solve the Schr\"odinger equation to calculate the fidelity
\be
 f = |\la\psi_{\rm ad}(t)|\psi(t)\ra|^2,
\ee
where $|\psi_{\rm ad}(t)\ra$ denotes the adiabatic state of $H_0(t)$
in equation~(\ref{ad2})
and $|\psi(t)\ra$ the solution of the Schr\"odinger equation 
with the Hamiltonian $H_0(t)+H_1(t)+\delta H(t)$.
We set the initial condition at $t=0$ to be the eigenstate of $S_1$ with 
the eigenvalue $+1$ and calculate the probability 
of observing the eigenstate of $S_2$ with the eigenvalue $+1$.
\begin{center}
\begin{figure}[t]
\begin{center}
\includegraphics[width=0.55\columnwidth]{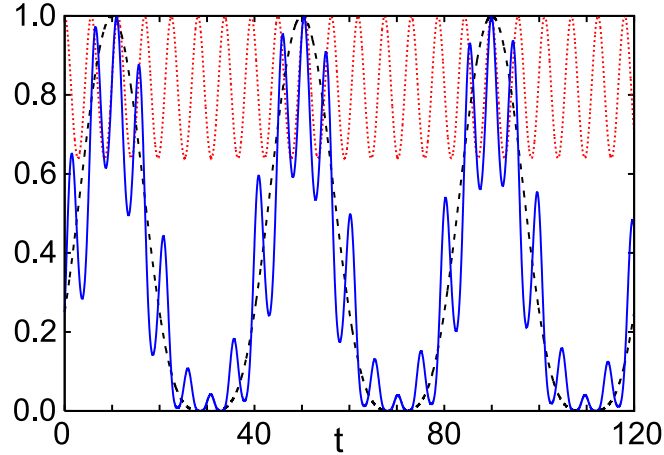}
\caption{Transitionless quantum driving with $S_3$-perturbation.
The red dotted line represents the fidelity and  
the blue solid line is the probability 
of observing the eigenstate of $S_2$ with the eigenvalue $+1$.
The black dashed line is the probability 
for the case of the ideal driving without perturbation.
}
\label{fig:s3}
\end{center}
\end{figure}
\end{center}
\begin{center}
\begin{figure}[t]
\begin{center}
\includegraphics[width=0.55\columnwidth]{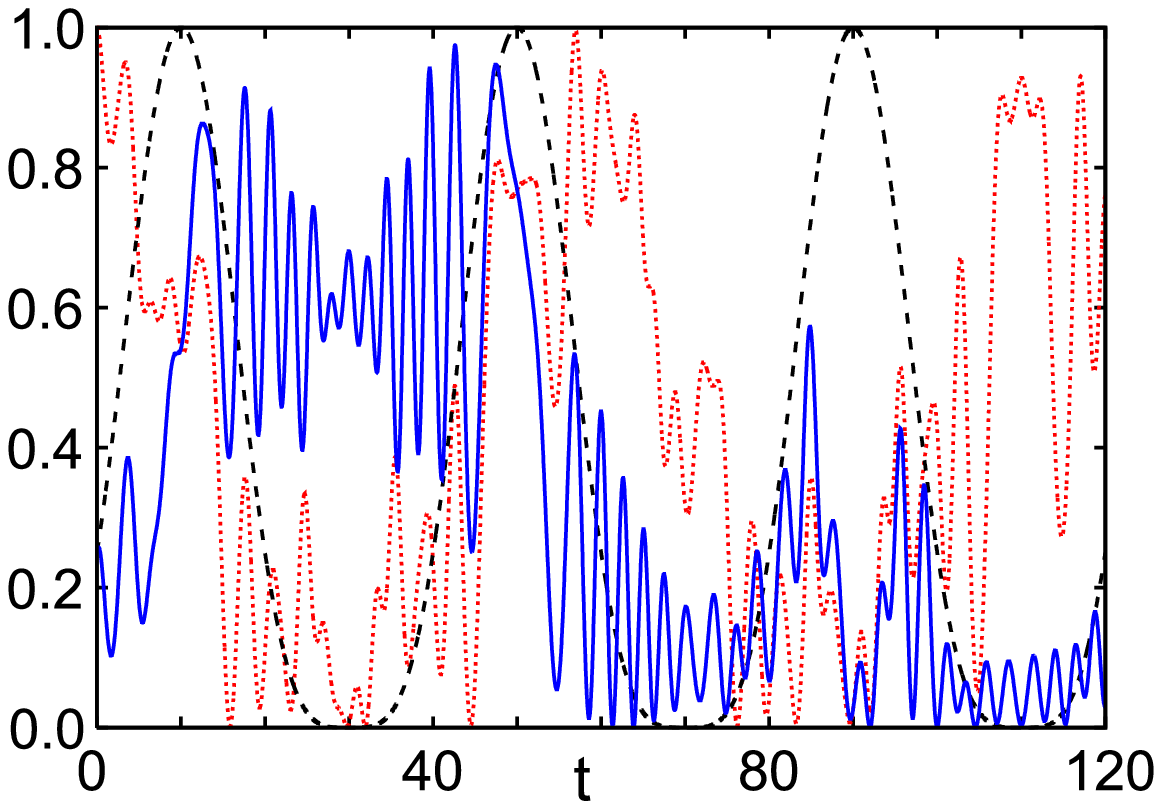}
\caption{Driving with $\lambda_4$-perturbation.}
\label{fig:l4}
\end{center}
\end{figure}
\end{center}
\begin{center}
\begin{figure}[t]
\begin{center}
\includegraphics[width=0.55\columnwidth]{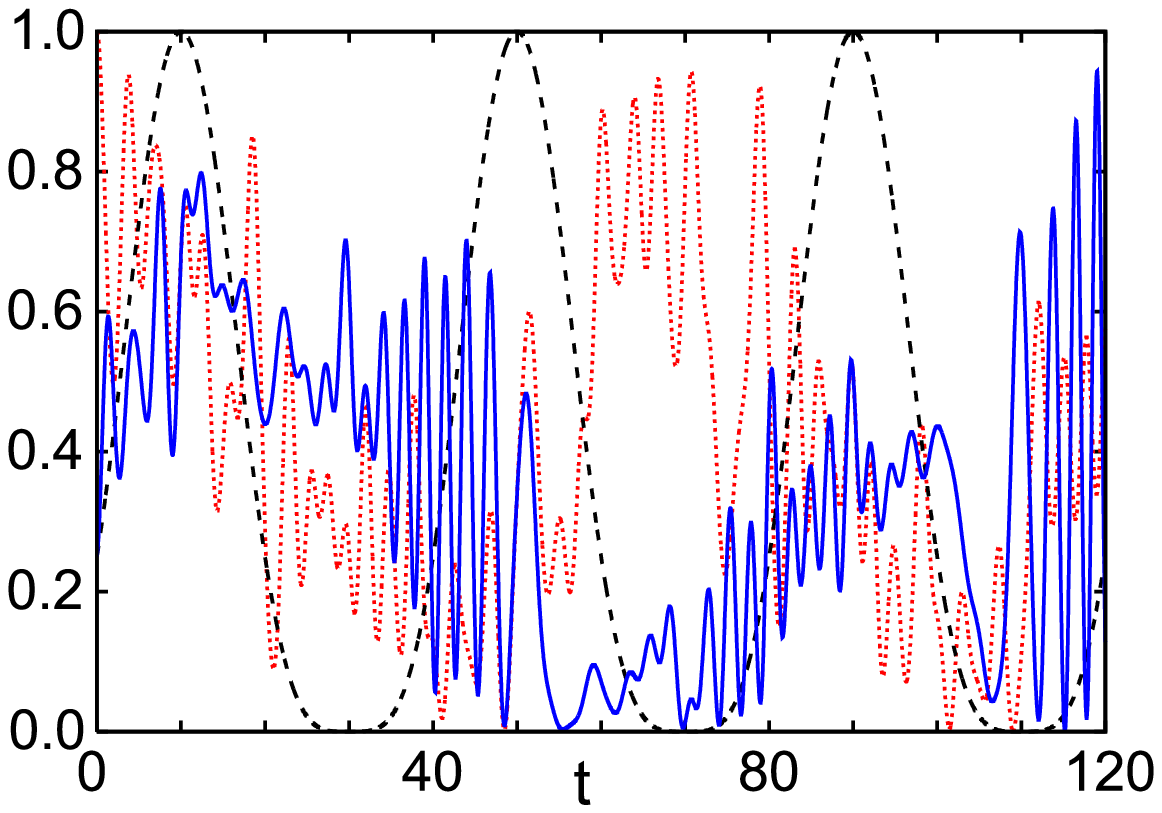}
\caption{Driving with $\lambda_5$-perturbation.}
\label{fig:l5}
\end{center}
\end{figure}
\end{center}
\begin{center}
\begin{figure}[t]
\begin{center}
\includegraphics[width=0.55\columnwidth]{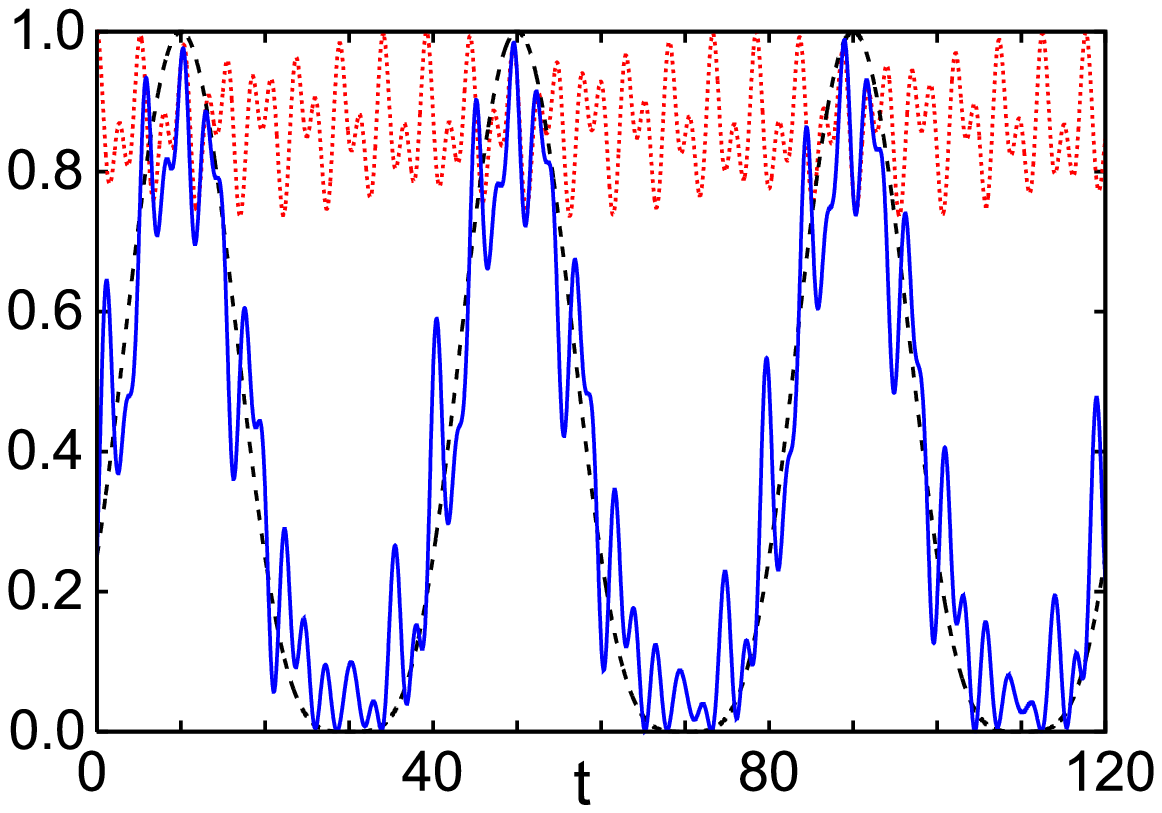}
\caption{Driving with $\lambda_8$-perturbation.}
\label{fig:l8}
\end{center}
\end{figure}
\end{center}

We set parameters $h_0=1$ and $\omega=\pi/20$.
As possible perturbations, we consider the following four cases:
\be
 \delta H(t) =  \delta h(t)\times\left\{
 \begin{array}{c}
 S_3 \\
 2\sqrt{\frac{2}{3}}\lambda_4 \\
 2\sqrt{\frac{2}{3}}\lambda_5 \\
 \frac{4}{3}\lambda_8
 \end{array}
 \right..
\ee
We set each coefficient so that 
the instability $I(t)$ calculated respectively as
\be
 I(t) = \frac{\delta h^2(t)}{\omega^2}\times\left\{
 \begin{array}{c}
 1 \\
 -\frac{2}{3}\left[1+\sin^2(2\omega t)\right] \\
 -\frac{2}{3}\left[1+\cos^2(2\omega t)\right] \\
 -1
 \end{array}
 \right.,
\ee
gives the same magnitude in average.
We set $\delta h(t)=0.5$.

The results are plotted in 
figures~\ref{fig:s3}, \ref{fig:l4}, \ref{fig:l5}, and \ref{fig:l8}.
We see in the stable case that the trajectory follows that 
of the ideal driving case while oscillating uniformly.
In unstable cases, we see more nonuniform behavior 
and large deviations for $\lambda_4$ and $\lambda_5$.
The deviation is relatively small for $\lambda_8$ 
but the oscillations are nonuniform compared with the stable case of $S_3$.
We note that $\lambda_8$ commutes with the driving term $S_3$ and 
does not disturb the system significantly.
Such a difference cannot be seen in the quantity $I(t)$.
We consider only the local instability of the solution of 
the QB equation.
Although it is interesting to see nonlinear effects, 
such an analysis is beyond the scope of this study.
The result here shows that the condition $I(t)>0$ can be 
one of the criteria for the stable driving.

%%%%%%%%%%%%%%%%%%%%%%%%%%%%%%%%%%%%%%%%%%%%%%%%%%%%%%%%%%%%%%%%%%%%%%%%%%%%
%%%%%%%%%%%%%%%%%%%%%%%%%%%%%%%%%%%%%%%%%%%%%%%%%%%%%%%%%%%%%%%%%%%%%%%%%%%%
\section{Summary}
\label{sec:summary}

We have studied the AP as a QB trajectory.
The QB equation shows that 
the Lewis--Riesenfeld invariant exists for the optimized trajectory
and the solution is interpreted as an AP.
As a practical situation, 
we closely discussed the case where 
the constraint part of the Hamiltonian gives the adiabatic state
so that one can control the system in a favorable way.
By considering several cases explicitly, 
we find that 
such a case is realized 
when the commutator of the basis operators 
between constraint parts $[X_a,X_b]$ does not belong to 
the constraint part.
This result will be a guiding principle 
in considering the ideal manipulation.

We note that we only considered the case where the constraint is 
given by the form (\ref{hconstraint}).
We can consider other types of constraints 
in the QB equation such as fixing $|\bm{h}(t)|$.
It is not clear what kind of the AP is realized in that case.
Such an analysis will be an interesting future problem.
In order to discuss the property of the AP more generally,  
it may be useful to consider a Lie-algebraic classification 
of the Lewis--Riesenfeld invariants as discussed in~\cite{GWFN}.

We also studied the stability of the solution and 
derived a general necessary condition for the stability. 
The result is confirmed in a three-level system.
Generally, the AP is stable against the variation of the operators 
which commutes with the counter-diabatic Hamiltonian $H_1(t)$
and unstable with operators anticommutes with $H_1(t)$. 

In this paper, we examined only the variation with respect to 
the Hamiltonian.
It is possible to consider the variation of the state.
Then, it can be possible to consider the stability under 
a change of the initial state for example.

Another possible study to be done is to interpret 
the fast-forward scaling~\cite{MN1, MN2, MN3} by using the QB.
In the fast-forward scaling, we utilize a reference state 
in fast-forwarding the evolution.
In such a formulation, the fidelity-optimized QB is 
expected to be suitable~\cite{KO}.
By clarifying the relations between various methods, 
we hope that we can understand nature of quantum fluctuations
in a deeper way, which will be useful for optimal coherent manipulations.

%%%%%%%%%%%%%%%%%%%%%%%%%%%%%%%%%%%%%%%%%%%%%%%%%%%%%%%%%%%%%%%%%%%%%%%%%%%%
\section*{Acknowledgments}

The author is grateful to S Masuda, Y Shikano and J Tsuda 
for useful discussions and comments. 

%%%%%%%%%%%%%%%%%%%%%%%%%%%%%%%%%%%%%%%%%%%%%%%%%%%%%%%%%%%%%%%%%%%%%%%%%%%%
%%%%%%%%%%%%%%%%%%%%%%%%%%%%%%%%%%%%%%%%%%%%%%%%%%%%%%%%%%%%%%%%%%%%%%%%%%%%

\end{document}